# Kondo effect in metallic glasses with non-Fermi liquid behavior

J. Q. Wang, W. H. Wang, and H. Y. Bai [*]

*Institute of Physics, Chinese Academy of Sciences, Beijing 100190, P. R. China*

ABSTRACT

By microalloying of Gd atoms with 4*f* electrons into CuZrAl or MgCuY glassy alloys, they display a variety of puzzling behaviors such as the characteristic of Kondo effect. The ground states of the Gd-alloyed systems are determined to have non-Fermi-liquid characteristics which derive from the strong structural disorder. The Kondo effect in these glassy alloys is attributed to the strong structural disorder. The coexistence of the Kondo effect and strong structural disorder has implications for the understanding the origin of the puzzling non-Fermi-liquid behavior.

The Kondo effect, which is known as the consequence of the strong exchange interaction between the localized spins and conduction electrons, is often observed in alloys that have local electrons well below the Fermi surface and that have very dilute local moments[1-2]. Systems with deeply localized moments often perform a magnetic order behavior at low temperatures that is dominated by direct interaction of local spins or by the so-called intersite Ruderman-Kittel-Kausuya-Yosida (RKKY) interaction if local spins are dilute. In crystalline metallic compounds with rare earth elements, the Kondo effect is usually observed in Ce, U, and Yb-based alloys[3-4]. Gd-contained compounds, which have too deeply-localized $f$ electrons, always behave magnetic ordering instead of Kondo effect. In this letter, we report that by microalloying of Gd atoms with $4f$ electrons into CuZr- and Mg-based glassy alloys[5], the characteristic of Kondo effect are observed in these BMGs, and the ground state of the systems is determined to be non-Fermi liquid (NFL) state. Such behavior can be modified by the concentration of the minor alloying Gd atoms and the degree of spatial disorder. The microalloying strategy opens a way for designing novel alloys with non-Fermi-liquid behavior.

Figure 1(a) shows the magnetic-field-driven behavior in $T$-dependent resistivity of Gd-alloyed $(CuZr)_{93-x}Al_7Gd_x$ (x= 0.5, 3, and 7) BMGs at low temperatures. The resistivity values have been normalized with the value at 300 K. The resistivity shows a weak negative linear $T$-dependent behavior at high temperatures, which is a common characteristic of glassy alloys[6]. The field-driven turn-up behavior at low-$T$ end is similar to that of crystalline $Ce_{1-x}La_xPb_3$ [7], which is a typical Kondo behavior. By subtracting the $T$-linear part at high temperatures as shown in Fig. 1(b), the contribution of local moments to the resistivity for x=3, 7 samples at low $T$ is obtained. A clear drop/shoulder around 35 K exist in the two resistivity curves, which is result from the crystalline electric field effect [7]. It can be seen that the position of the minima and its depth relative to the value at $T = 0$ K depend on the concentration of local moments, which is also a typical character of Kondo effect [8]. The character temperature $T_K$ of the Kondo effect in such a strong structure-disordered system should have a distribution rather than an exact value [10].

Nevertheless, according to $R(T_K) \sim R_{max}/2$ [8], an effective $T_K$ is determined to be 1.10(8) K and 2.35(9) K for x=3 and 7 samples, respectively. The determined $T_K$ is in good consistence with the specific heat results shown below.

In semi-log plot of Fig. 1(c), the $T$-dependent resistivity for x=3 sample changes from an $\ln T$ dependence to the form of $R \sim T^{2-\lambda}$ ($\lambda$=1.26) with decreasing $T$. According to Landau's Fermi liquid theory, the resistivity should get to a form with $R \sim T^2$ at low temperatures [3]. However, the $T$-dependent resistivity of the glass at low temperatures deviates far from $R \sim T^2$ as shown in Fig. 1(c). This is a typical non-Fermi-liquid (NFL) behavior [3, 10-11]. In such a heavily structural disordered glassy system, it is thought that the NFL behavior derives from the strong atomic disordered structure. The Kondo effect and NFL behavior are also observed coexist in another heavily structural disordered Gd-alloyed $Mg_{65}Cu_{25}Y_5Gd_5$ glassy system [see Fig. 1(d)]. Its $T$-dependent resistivity follows a form of $R \sim T^{1.5}$ when $T < 7$ K.

Figure 2 shows susceptibility data for x=3 sample in the form of $1/\chi$ vs. $T$. The inset shows the data in form of $\chi$ vs. $T$. At high temperatures, the susceptibility of the metallic glass follows the Curie-Weiss behavior $\chi = \dfrac{A}{T-\theta_C}$, with $A$=8.19(8) emu.K/mol-Gd.Oe and $\theta_C$=21.5(5) K. The effective magneton $\mu_{eff}$ of Gd atoms is determined to be 8.04(9) $\mu_B$ by $A = \mu_0 \mu_{eff}^2/3k_B$, which is close to the theoretical value of 7.94$\mu_B$. The deviation from Curie-Weiss law at low temperatures derives from the antiferromagnetic interaction between $f$ and conduction electrons, namely the Kondo compensating/screening effect, which would weaken the effective magneton of Gd atoms and decrease the susceptibility [9]. As shown in the inset, the susceptibility for x=3 sample at low magnetic field (0.1 Tesla) shows a $\chi \sim T^{1+\lambda}$ ($\lambda$=0.89, the cyan-colored line) behavior over a decade T-range, and this is a typical characteristic of NFL behavior [4, 10].

The specific heats $C_p$ for Gd-alloyed samples of $(CuZr)_{93-x}Al_7Gd_x$ (x= 0.5, 3, 5, 7,10) at low temperatures were measured. The $C_p$ of a Y-alloyed $(CuZr)_{90}Al_7Y_3$ BMGs that was also measured and shown in Fig. 3(a). Y has similar valences and ionic radii to Gd, but without 4$f$ electrons, the corresponding CuZrAlY BMG was used as a reference for obtaining the specific heat of 4$f$ electron contribution. The $C_p$ of the Y-alloyed sample well fits the Debye model of $C_p=\gamma T+\beta T^3$ with $\gamma$=2.40(12) mJ/mol-atom.K$^2$ and $\beta$=0.15(2) mJ/mol-atom.K$^4$. And the value of $\gamma$ is close to that of many typical metallic glasses[12]. Comparatively, due to 4$f$ electrons, $C_p$ of the Gd-alloyed samples deviates from the Debye model with a Kondo peak at the low-$T$ end. According to the single-ion Kondo model [13], the peak temperature $T_{max}$ can be determined with $T_{max}$~0.7$T_K$. For instance, the $T_{max}$ in experimental data are about 0.68(7) K and 1.65(9) K, and these result in $T_K$ = 0.97(10) and 2.36(13) K for x=3 and 7 glasses, respectively. The results are in good agreement with that determined from the resistivity measurements. The $C_p$ data of a series of samples show that the contribution of the Kondo effect to $C_p$ tends to increase with the increase of Gd up to 5 at.%, while the contribution decreases with further increase of Gd concentration due to the competition with the strengthened RKKY interaction between the local moments. Figure 3(b) shows $C/T$ vs. $T$ in semi-log plot for various Gd-alloyed CuZr-based BMGs obtained by subtracting the $C_p$ data for the corresponding Y-alloyed BMGs, the value of $C/T$ has also been normalized to Gd atoms. It can be seen that the $C_p$ also show characteristic of NFL at low T, with $C/T \sim T^{-\alpha}$ [4] across almost a decade $T$ range. The fitting parameters $\alpha$ are determined to be 1.67, 1.02, 0.11, 0.07 and 0.49 for x=0.5, 3, 5, 7, 10 samples, respectively.

Both the Kondo effect and RKKY interaction correlate with a $g$ factor in the following forms [3]. The binding energy of Kondo effect can be identified with $k_B T_K \propto \exp\left(-\dfrac{1}{g}\right)$, while the strength of RKKY interaction can be characterized by $k_B T_{RKKY} \propto g^2$, with $g=N(E_F)J$, and $J \cong \dfrac{V_{k,f}^2}{E_F-E_f}$, where $k_B$ is the Boltzman constant, $J$ is coupling constant, $V_{k,f}$ is the hybridization matrix between $f$ state and conduction electrons, $E_F$ is the energy of Fermi level, $E_f$ is the energy of $f$ state level, $N(E_F)$ is the conduction-band density of states at $E_F$. The Kondo effect deriving from the antiferromagnetic interaction between the local moment and free electrons competes with the RKKY interaction. For systems with big $V_{k,f}$ and small $E_F - E_f$, the Kondo behavior dominates at low T; oppositely, the RKKY interaction dominates. In normal rare earth compounds, the *f* electrons are localized and the interactions with the conduction electrons can be described by the Heisenberg s-*f* exchange interaction model with a ferromagnetic coupling [8]. The *f* levels lie too far below the Fermi level, for the indirect exchange by virtual hopping into the conduction band to be important, and the direct exchange term of the Coulomb interaction between the *f* and conduction electron, which is ferromagnetic, dominates. This interaction then leads to an intersite exchange interaction between the localized *f* moments mediated by the conduction electrons, the RKKY interaction, which may be antiferromagnetic, ferromagnetic, or both because it extends beyond nearest neighbor pairs [8, 14]. The Gd atom has the deepest localized *f* electrons among the rare earth elements[15], and is thought to be impossible to perform Kondo interaction[16]. The Gd-alloyed compounds often perform the RKKY-interaction-determined long-range magnetic order. The observation of on-site Kondo effect in the Gd-alloyed metallic glasses result from the strong structural disorder in the glasses that results the broadening of *f* energy band, and then some local *f* electrons of Gd atoms are much close to the Fermi energy level and makes the interaction between localized *f* electrons and conduction electrons possible. The 4*f*-electrons form a high DOS energy band below the conduction band, with an energy-gap ($E_g$) existing between the two bands. The 4*f*-electrons, like the particles in an energy trap, have a possibility ($\propto E_g^{-2/3}$) to get out according to the quantum mechanics theory[17]. In BMG with strong structural disorder, heavy-band-tails are introduced into the conduction band and many electrons are bounded in these local states induced by structural disorder[18]. These make $E_g$ smaller, and the exchange interaction between the two kinds of electrons become easier to happen. Such hybridization between 4*f* and conduction electrons is vital in Kondo effect [19].

To study the role of the structural disorder in the observed Kondo behavior, the degree of the structural disorder of $(CuZr)_{90}Al_7Gd_3$ BMG was modified by isothermal annealing under high vacuum at 663 K (23 K below the glass transition temperature $T_g$=686 K) for 20, 100 hrs. Annealing below $T_g$ can make the metastable metallic glasses relax structurally to a more ordered state and modify the degree of the structural disorder of a glass [20]. The fully-crystallized BMG was also obtained by annealing the BMG at 793 K (43 K above the crystallization temperature, $T_x$=750 K) for 3 hrs. Figure 4 shows the $C_p$ for the as-cast and annealed $(CuZr)_{90}Al_7Gd_3$ BMGs. Obviously, the annealing induces obvious frustration of the Kondo effect contribution in the low-$T$ specific heat anomaly. This denotes that structural disorder strengthens the Kondo effect by suppressing the long-range RKKY interaction. The crystallized state with mixed crystalline phases has also heavily structural disorder especially in the boundaries between phases, and this residual disorder should be important season for the remained Kondo interaction in the crystallized BMG. Schottky anomaly, which derives from the energy level splitting induced by crystalline electric field (CEF) effect, sometimes exists a system with local magnetic moments, and could give a similar contribution to low-$T$ specific heat. The increase of structural order by annealing would strengthen the CEF, and the Schottky anomaly contribution to the $C_p$ should be strengthened. However, the annealing $C_p$ measurements show an opposite tendency. Therefore, together with the observation of typical Kondo characters in $T$-dependent resistivity and susceptibility measurements, the Schottky anomaly should be very weak and negligible compared with the Kondo effect in the glasses.

The unique combination of the NFL behavior, the strongly-disordered structure, continuously-tunable $f$-atom concentration, and the liquid-like spatial and direction uniformity makes the BMGs an ideal system to study the correlations between the Kondo effect and the concentration of the $f$-atoms. The strategy could be applicable to the development of a series of glassy materials with strongly correlated electronic states and open up a research area of both fundamental and applied importance. The study of disorder-driven special properties that derive from the minor atoms with local moments will also helpful for understanding the local atomic structures in metallic glasses.

**Acknowledgements:** The discussion with Prof. Y. Wu, M.W. Chen is appreciated. Financial support is from the NSF of China (Nr. 50671117) and MOST 973 (No. 2007CB613904).

**Figure Captions**

Figure 1. (Color online) (a) $T$-dependent resistivity of $(CuZr)_{93-x}Al_7Gd_x$, x= 0.5, 3, and 7, under different magnetic fields. (b) The local moments contribution to resistivity for x=3 and 7 samples are got by subtracting the $T$-linear part of high-$T$ region. (c) Resistivity of x=3 sample gets to a finite value after an $R \sim \ln T$ region when cooled down, in $R \sim T^{2-\lambda}$ ($\lambda$=1.26) form rather than $R \sim T^2$. (d) Kondo behavior is also observed in Gd-alloyed MgCuY BMG. The green line shows the fitting data of $R \sim T^{1.5}$, denoting a NFL behavior.

Figure 2. (Color online) The susceptibility under various magnetic fields of x=3 sample in the form of $1/\chi$ vs. $T$. The inset shows the corresponding data in form of $\chi$ vs. $T$, with the fitting data $\chi \sim T^{-1+\lambda}$ ($\lambda$=0.89) for low field data.

Figure 3. (Color online) (a) Specific heat of $(CuZr)_{93-x}Al_7Gd_x$ (x=0.5, 3, 5, 7, 10) BMGs. The filled symbols are the $C_p$ of the BMGs with a peak-like deviation from Debye model. The dark solid line is the fitting for $(CuZr)_{90}Al_7Y_3$ according to Debye model, $C_p=\gamma T+\beta T^3$, with $\gamma$=2.40(12) mJ/mol-atom.K$^2$, $\beta$=0.190(2) mJ/mol-atom.K$^4$. (b) The $C_p/T$ vs. ln$T$. The solid lines show the fitting of $C_p/T \sim T^{-\alpha}$, with $\alpha$=1.67, 1.02, 0.11, 0.07 and 0.49 for x=0.5, 3, 5, 7, 10 samples, respectively, which is a typical character of NFL behavior.

Figure 4. (Color online) The annealing effects on $T$-dependent $C_p$ of $(CuZr)_{90}Al_7Gd_3$ sample with different annealing times, $t_a$= 0, 20, 100 hrs, at 663 K ($< T_g$). The data of the crystallized sample is also shown.

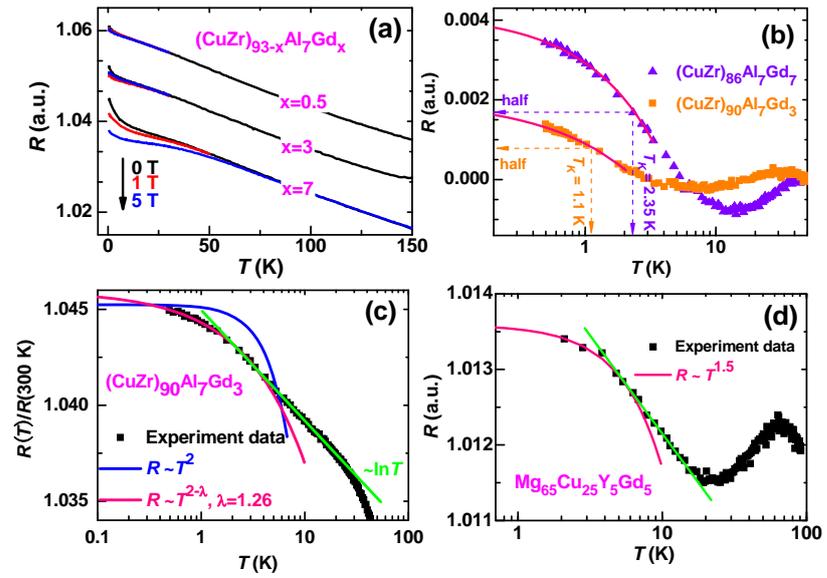

Figure 1, J. Q. Wang, *et al*

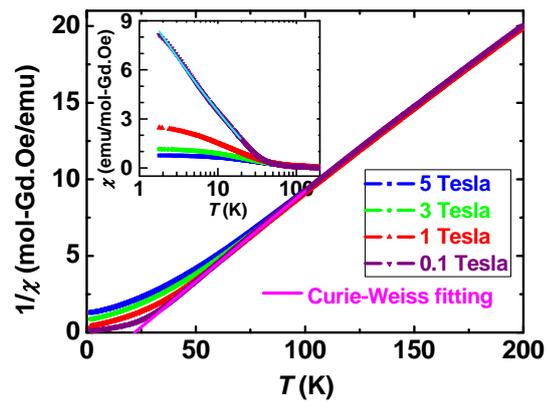

Figure 2, J. Q. Wang, *et al*

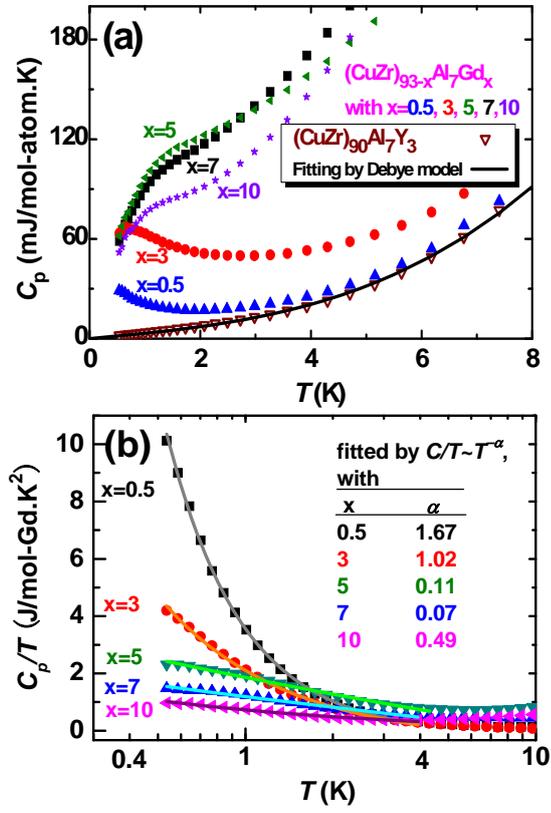

Figure 3, J. Q. Wang, *et al*

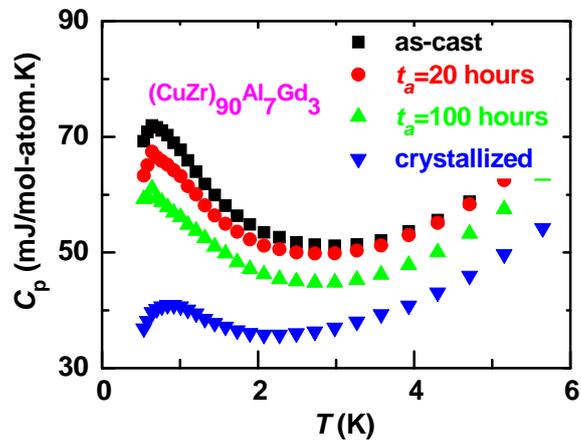

Figure 4, J. Q. Wang, *et al*